\documentclass[prl,a4paper,twocolumn,footinbib,showpacs,floatfix]{revtex4}

    \usepackage[latin1]{inputenc}
    \usepackage{amssymb}
    \usepackage{amsmath}
    \usepackage{amsbsy}
    \usepackage{bm}
    \usepackage{txfonts}
    \usepackage{graphicx}

    \newcommand{\I}{\mathrm{i}}
    \newcommand{\E}{\mathrm{e}}

\begin{document}
\title{Vortex sheet dynamics and turbulence}

\author{Malek Abid}\email{abid@irphe.univ-mrs.fr}
\author{Alberto Verga}\email{verga@irphe.univ-mrs.fr}
   \affiliation{%
   Institut de Recherche sur les Phénomènes Hors \'Equilibre,
   UMR 6594, CNRS, Université de Provence, Marseille, France}%
   \altaffiliation{%
	IRPHE, 49, rue F. Joliot-Curie, BP 146, 13384 Marseille, France
   }

\date{\today}

\begin{abstract}
The nonlinear evolution of a vortex sheet driven by the Kelvin--Helmholtz instability is characterized by the formation of a spiral possessing complex stretching and intensity patterns. We show that the power energy spectrum of a single two-dimensional vortex sheet tends to the usual fluid turbulent spectrum, with an exponent of $-3$. Using numerical simulations and asymptotic methods, we demonstrate the relation between this power law and the singularities in the geometry and vorticity distribution of the sheet.
\end{abstract}

\pacs{47.15.Ki, 47.32.C-, 47.27.E-}

\maketitle

\begin{figure*}
    \centering
    \includegraphics[width=0.3\textwidth]{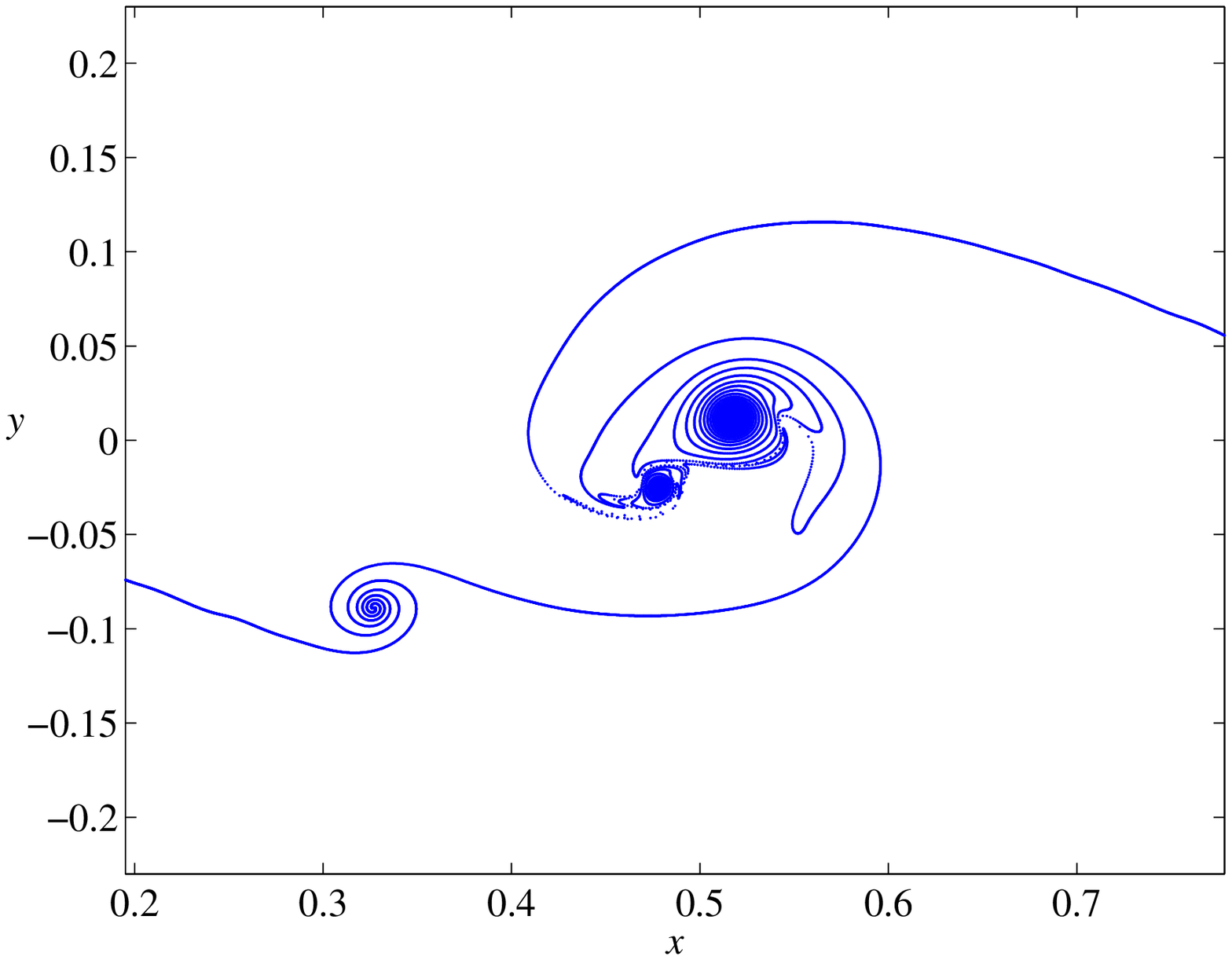}(a)$\;$%
    \includegraphics[width=0.28\textwidth]{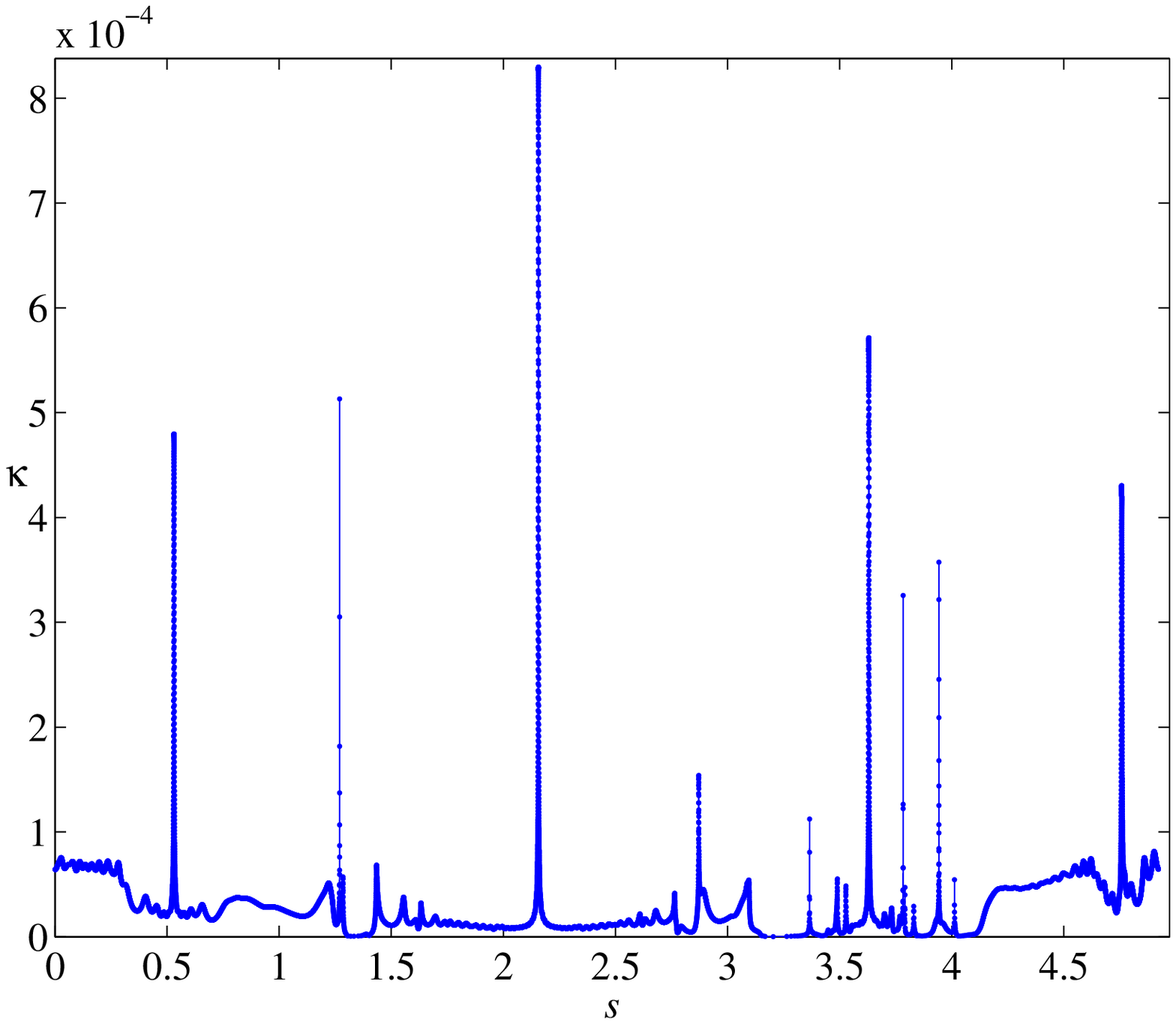}(b)$\;$%
    \includegraphics[width=0.32\textwidth]{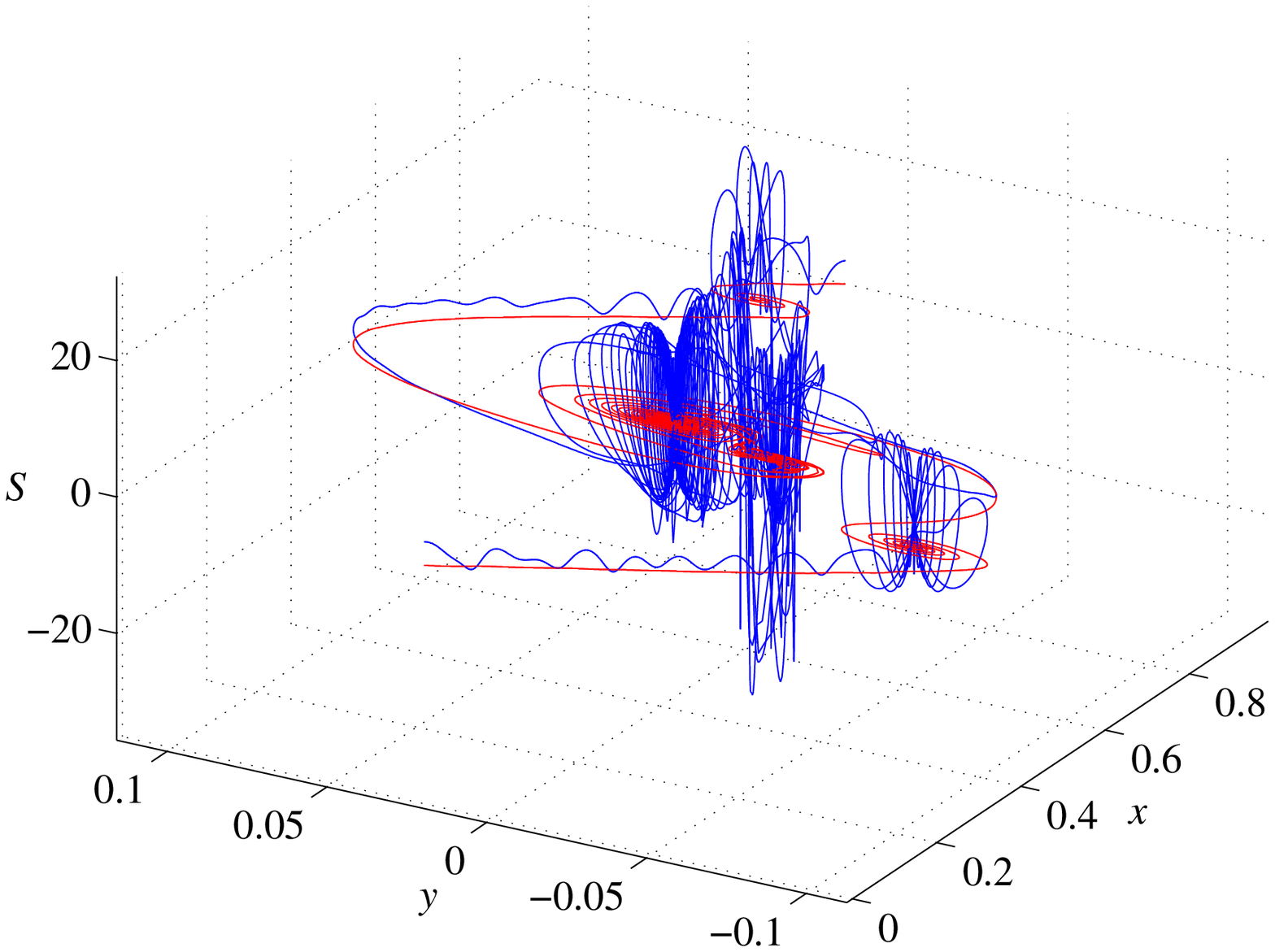}(c)
    \caption{\label{fig_xy} (Color online.) 
Spiral formed at the nonlinear stage  of the Kelvin--Helmholtz instability ($N=16384$, $\delta=0.024$, $t=0.93$). (a) Snapshot of the vortex sheet; (b) the intensity $\kappa(s,t)$ concentrates in localized regions; (c) the strain rate $S(s,t)$ (in blue) shows a complex structure of alternating compression and stretching zones (in red, the vortex sheet).
    }
\end{figure*}

In 1941, Kolmogorov formulated a theory of fluid turbulence, based on the idea of statistical similarity \cite{Kolmogorov41}. In parallel, Onsager proposed that a singular enough velocity field could explain dissipation without the ``final assistance of viscosity'' \cite{Onsager45,Onsager49}. (A thorough review on Kolmogorov theory can be found in the book by U. Frisch \cite{Frisch95}. Onsager's work on turbulence is accounted for in Ref.~\cite{Eyink06}.) These two complementary approaches, the one statistical, the other dynamical, can be taken as constitutive of further theoretical developments on turbulence. One difficult problem, whose solution would contribute to understanding the physics of turbulent flows, is to establish the relation existing between dynamics and statistics, in particular flows. Following the theoretical ideas first introduced by Townsend \cite{Townsend51}, Lundgren proposed a relation of this type in his model of intermittent structures at high Reynolds numbers \cite{Lundgren82,Pullin98}. Lundgren demonstrated that solutions of the Navier--Stokes equation in the form of slender spiral vortices yield Kolmogorov spectrum, through a mechanism of enstrophy intensification and creation of fine scales by differential rotation and stretching. In two dimensional flows, Gilbert \cite{Gilbert88} studied the wind-up of a vortex patch by a strong point vortex, and found that an energy spectrum $E(k)\sim k^{-\gamma}$ with $3<\gamma<4$, is obtained from the fractal structure of the spiral separating potential and vortical regions. The goal of this Letter is to study the relation between geometry, vorticity distribution and spectral properties of the flow generated by the winding-up of a shear layer subject to the Kelvin--Helmholtz instability. 

In two dimensions, and in the limit of vanishing viscosity, a shear layer becomes a line of velocity discontinuity. The Euler flow associated with a vortex sheet reduces to the Hamiltonian dynamics given by the Birkhoff--Rott equation \cite{Sulem81,Saffman92},
\begin{equation}\label{BRper} 
	\frac{\partial}{\partial t}\bar{Z}(s,t)=
    \frac{1}{2\pi\I} \fint ds' \kappa(s',t) 
		K\bigg(Z(s,t)-Z(s',t)\bigg)\,.
\end{equation}
In Eq.~(\ref{BRper}) the vortex sheet is represented by the parametric expression of the curve $z=Z(s,t)$, where $z=x+\I y$ is a complex coordinate and $s$ is the arclength, $ds=|dZ|$. The Biot--Savart kernel is defined by $K(z)=1/z$ or $K(z)=\cot(\pi z)/2$ in the case of periodic boundary conditions in the $x$-direction. In (\ref{BRper}) $\kappa=\Gamma_s\equiv\partial \Gamma/\partial s$ (subscript $s$ denotes derivative) is the intensity of the sheet, where the circulation $\Gamma=\Gamma(s,t)$ is a Lagrangian variable satisfying $d\Gamma/dt=0$; $\fint$ denotes the Cauchy principal value integral, and the over bar denotes the complex conjugate. A sinusoidal perturbation of a straight vortex sheet evolves towards a singularity in finite time, as demonstrated by Moore in 1979 \cite{Moore79,Cowley99}, probably related to the incipient rolling-up of the sheet into a spiral. In order to integrate (\ref{BRper}) in time, it is then necessary to regularize the kernel. A simple method consists in cutting the vortex interaction at a distance $\delta$, therefore replacing $K\sim1/z$ by $K_{\delta}=\bar{z}/(|z|^2+\delta^2)$ \cite{Krasny86a}. Solutions of the regularized Birkhoff--Rott equation, with $K$ replaced by $K_{\delta}$, tend to solutions of the Euler equation in the limit $\delta\rightarrow0$ \cite{Liu95}. It is worth noting that the dynamics of the desingularized (\ref{BRper}) remains Hamiltonian. As a consequence, this model is particularly well suited to investigate the dynamics of the turbulent cascade in one of the most basic flow structures, the Kelvin--Helmholtz unstable vortex sheet.

We used the blob vortex method of Krasny \cite{Krasny86a} to compute the dynamics of a shear layer in a periodic domain of length $L$. We take units such that $L=1$ and the circulation is in the interval $\Gamma\in[0,1]$ (or the unperturbed velocity discontinuity equal to one). The initial perturbation, as for example the superposition of small amplitude modes, evolves towards the formation of a system of spirals winding around the main one. These structures are driven by a cascade of secondary Kelvin--Helmholtz instabilities, as shown in Ref.~\cite{Abid02}. Figure~\ref{fig_xy} presents the vortex sheet formed from an initial condition of two modes with wavelengths $1$ and $1/4$, together with the strain rate $S(s,t)$ and the intensity $\kappa(s,t)$. The strain rate is defined by $S=(1/\Delta s)d\Delta s/dt$, where $\Delta s\rightarrow0$, is the arclength separation of two neighboring points on the vortex sheet. Diminishing the value of $\delta$, the relevant physical limit is $\delta\rightarrow0$, results in a drastic increasing of small scale features of the sheet winding \cite{Krasny91} (see also \cite{Majda02}). We study numerically the generation of the small scales resulting from the short wave instabilities, in the presence of nontrivial stretching and vorticity concentration effects. Indeed, arbitrarily small scale perturbations might grow, depending for instance on the local strain rate \cite{Abid02}, and regions of strong localized vortex intensity develop as observed in Fig.~\ref{fig_xy}(b). The origin of these small scale features, can be traced back to the highly oscillatory behavior of the vortex sheet stretching, whose dynamics is essentially Lagrangian and nonlocal (Fig.~\ref{fig_xy}(c)). The presence of strong velocity gradients (in the limit $\delta\rightarrow0$) possessing a wide range of space and time scales, responsible for the appearance of an increasingly complex velocity field, can be considered to be the root of a turbulent flow. We found indeed, that the flow created by this single shear layer becomes \emph{turbulent}, in the sense that it is characterized by a well defined power law energy spectrum $E_k\sim k^{-\gamma}$ (for $k\ll 2\pi/\delta$, and $3<\gamma<4$), in agreement with full Navier--Stokes simulations with random initial vorticity field \cite{Brachet88,Tabeling02}.

It is then natural to look at the energy spectrum, $E_k=E(k)$ of the sheet. However, in order to avoid the arbitrariness in the origin of coordinates we define a one-dimensional spectrum from the $y$-direction average of the velocity $\bm{v}(x,y)$ correlation function $c(\xi)$ , 
\begin{equation}\label{Ek}
E_k  \equiv E(k) = \frac{1}{2}\mathcal{F}_k [c(\xi )] =
\frac{1}{2}\int\limits_0^{1 } d\xi c(\xi )e^{ -
ik\xi } \,,
\end{equation}
where $\mathcal{F}_k$ is the Fourier transform, and
\begin{equation}\label{cxi}
c(\xi ) = \frac{1}{2l}\int\limits_{-l}^l dy\int\limits_0^1 dx
\bm{v}(x + \xi ,y)\cdot\bm{v}(x,y) \,.
\end{equation}
(The averaging length $l$ is chosen munch larger than the spiral size.) This definition, similar to the one proposed by Saffman and others \cite{Saffman71,Gilbert88,Moffatt93}, satisfies the condition that the conserved total energy $E$, is given by $\int_0^{\infty} dkE(k) = \int_0^{\infty}dk|\bm{v}_k|^2= E$, where, $\bm{v}_k=\mathcal{F}_k[\bm{v}]$. We measured the spectrum (\ref{Ek}) of the vortex sheet shown in Fig.~\ref{fig_xy}, and found that for $k\ll 2\pi/\delta$, it is well fitted by a power law  $E_k\sim k^{-2.94}$ close to the law observed in developed decaying two-dimensional turbulence \cite{Brachet88}. In fact, the actual value of the exponent follows the evolution of the initially regular vortex sheet. The value $\gamma\approx3$ is reached after the development of the secondary structures (see the insert in Fig.~\ref{fig_xy}). At the early stages of the winding, intermediate values $3<\gamma<4$ are measured, with a low wavenumber range dominated by $\gamma\approx3$, and a large wavenumber range having a steeper spectrum. The $\gamma\approx3$ range of $k$ increases with time, and at $t=0.386$ it extends almost over all the available range ($0<k<2\pi/\delta$). In order to obtain a significant range in the turbulent spectrum we proceeded for the first time to a quadruple precision computation (33 significant digits), using $\delta=0.01$ and $N=2^{15}$ vortices.

\begin{figure}
    \centering
    \includegraphics[width=0.3\textwidth]{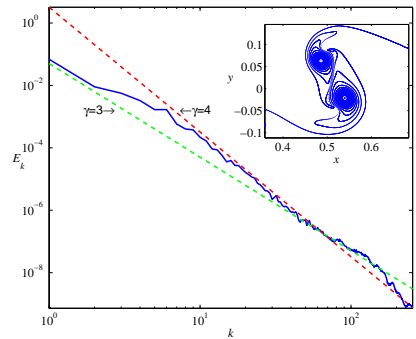}
    \caption{\label{fig_Ek}(Color online.)
The energy spectrum (\protect\ref{Ek}) of the flow associated with the vortex sheet shown in the insert  ($N=2^{15}$, $\delta=0.01$, $t=0.386$, quadruple precision computation).
    }
\end{figure}

As it is well known, the superposition of velocity discontinuities yields a spectrum in $E_k \sim k^{-2}$ \cite{Townsend51}, while a superposition of vorticity discontinuities yields $E_k \sim k^{-4}$ \cite{Saffman71}, the classical enstrophy cascade picture gives $E_k \sim k^{-3}$ \cite{Batchelor69}. It would be interesting, as suggested by Moffatt \cite{Moffatt84}, to establish a relationship between the fractal structure of the rolled shear layer and the observed energy spectrum, characterized by an intermediate value of the exponent. (It is worth mentioning that for finite $\delta$ the velocity field is continuous.) The fractal dimension $d_F$ of the spiral in Fig.~\ref{fig_xy} is found to be $d_F\simeq 1.44$. This value can be compared to $d_F=1.33$ measured in Ref.~\cite{Angilella99}, for a single Kelvin--Helmholtz spiral and for a large value of the smoothing parameter $\delta=0.25$. Assuming that the formula $\gamma=-5+d_F$ relating the energy exponent to the fractal dimension, derived by Gilbert \cite{Gilbert88} in his study of a spiral vortex patch, holds in the present case, we obtain a value $\gamma\approx 3.56$, somewhat larger than the observed one. The difference may be attributed to the nonuniformity in the distribution of the vorticity.

\begin{figure*}
    \centering
    \includegraphics[width=0.3\textwidth]{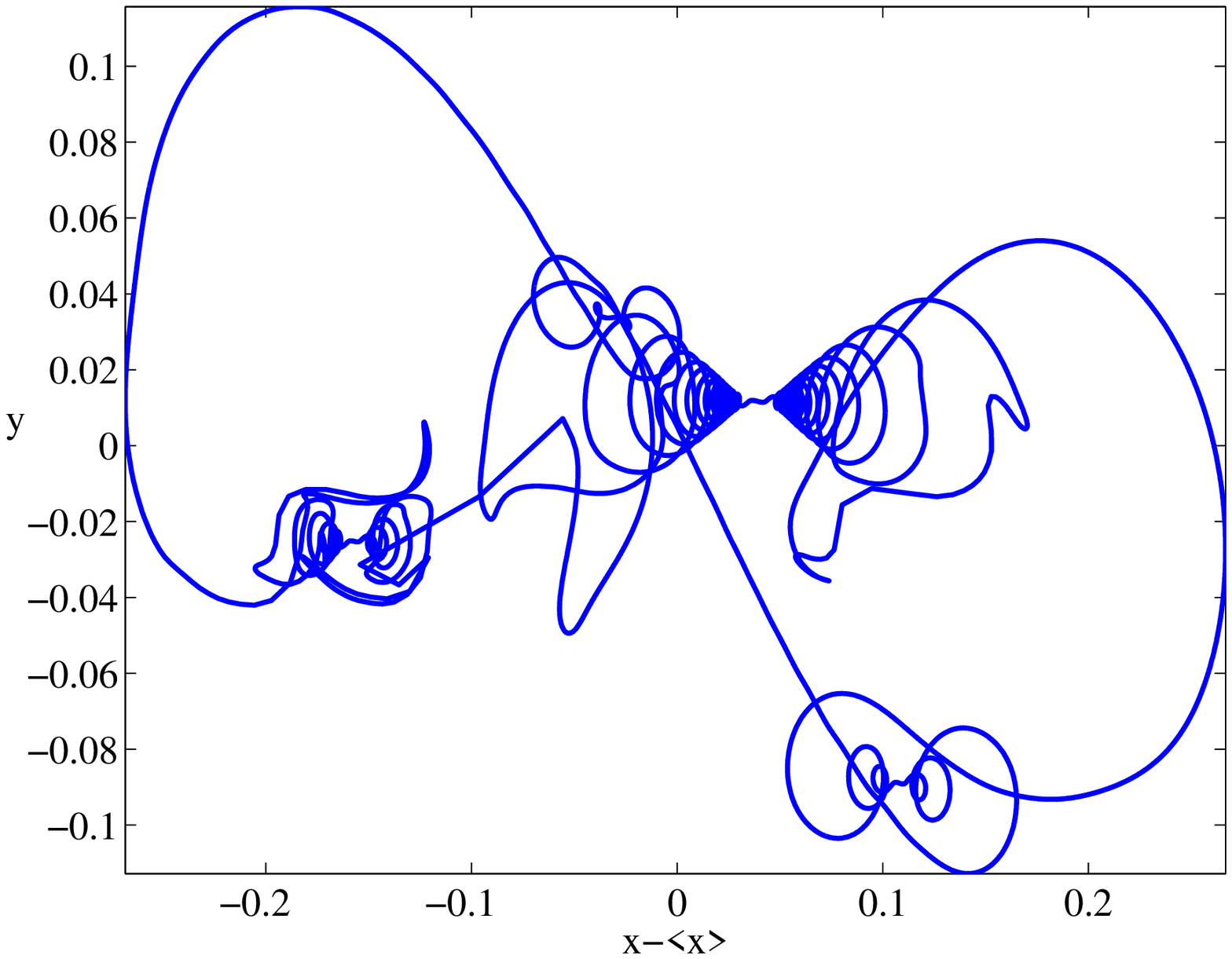}(a)$\;$
    \includegraphics[width=0.3\textwidth]{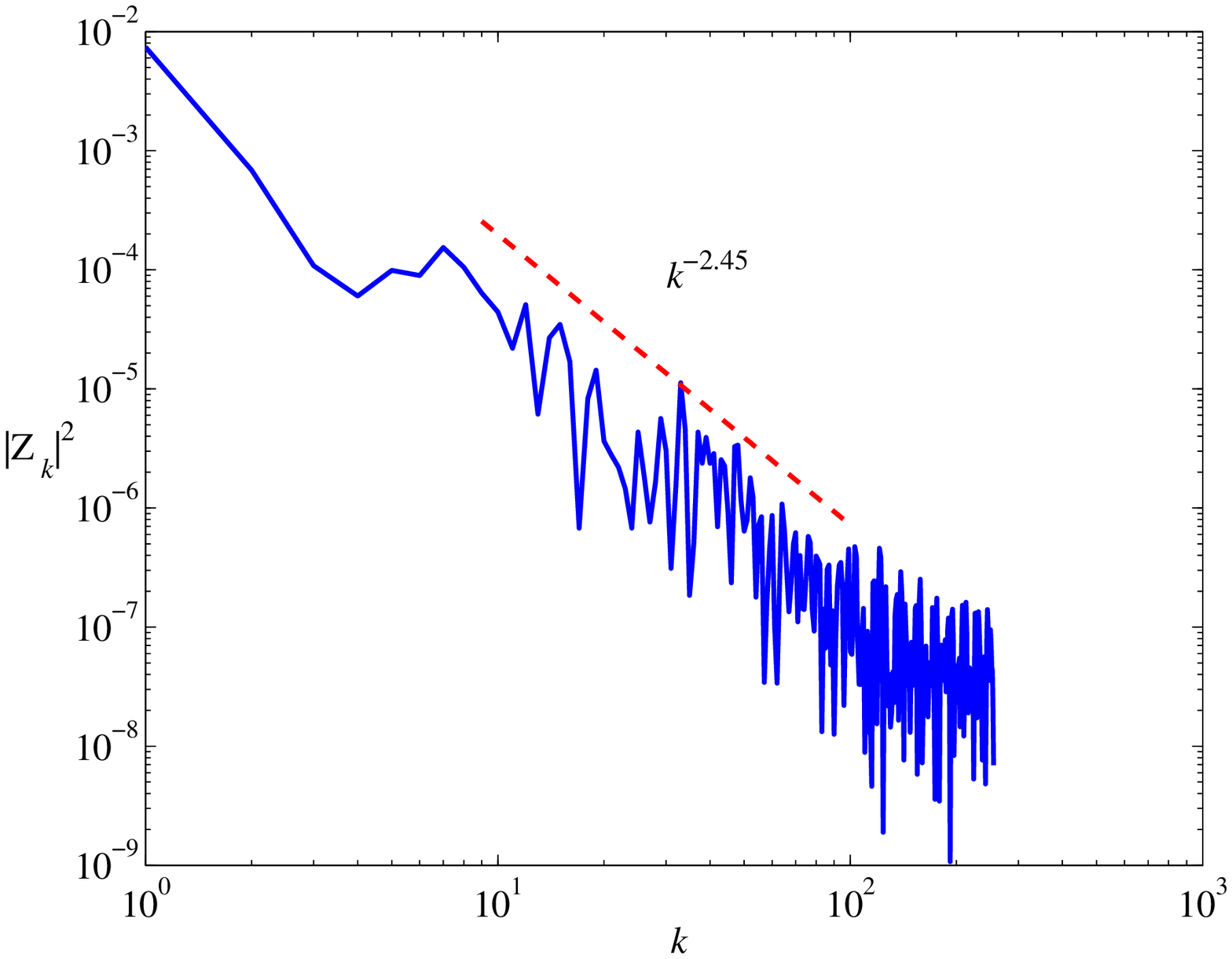}(b)$\;$
    \includegraphics[width=0.3\textwidth]{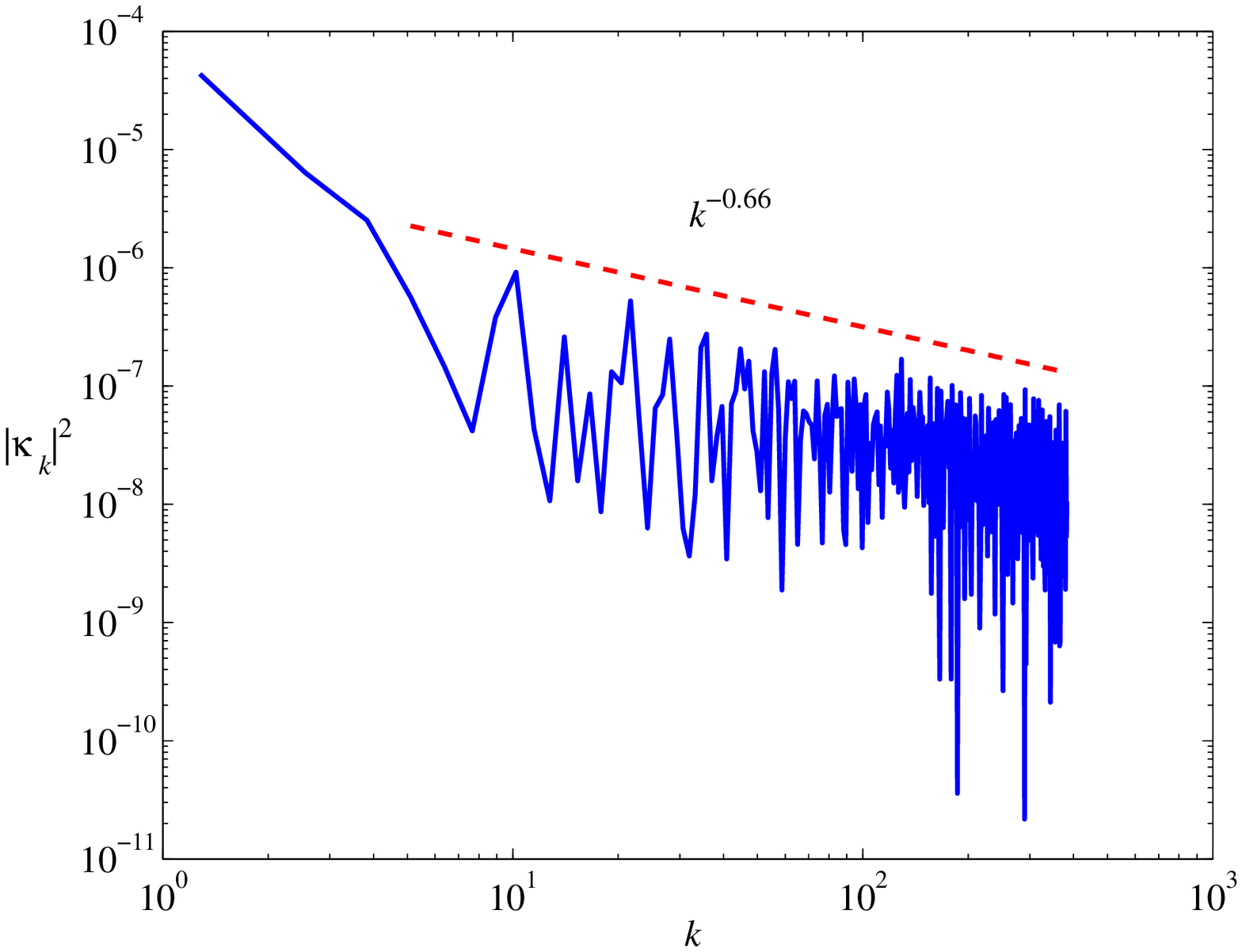}(c)
    \caption{\label{fig_Z233}(Color online.)
Shape and spectrum of the vortex sheet. (a) The spiral shown in Fig.~\protect\ref{fig_xy} after periodization in the $x$ direction; (b) power spectrum of $Z(s)$, with exponent $2\alpha+2=2.45$; (c) power spectrum of $\kappa(s)$, with exponent $2\beta+2=0.66$.    
    }
\end{figure*}

It is then tempting to conjecture that the characteristic exponent of the spectrum $\gamma$ reflects, not only the fractal spiral structure of the sheets, but also the existence of singularities in both the shape and the intensity of the vortex sheet. Remark that the existence of such type of singularities is not related to eventual singularities in Euler flows, the velocity field outside the sheet is always regular in our case. Indeed, we can use the Biot--Savart equation for the velocity field to relate $\gamma$ with the characteristic exponents $\alpha$ and $\beta$ of the shape $Z(s)\sim s^\alpha$ and intensity $\kappa(s)\sim s^\beta$ singularities, respectively:
\begin{equation}\label{BS}
    v_k \sim \int ds \int dx\frac{s^\beta\E^{- \I kx} }{x-s^\alpha } 
		\sim k^{-(\beta+1)/\alpha}\,.
\end{equation}
Therefore, $E_k  = |v_k |^2  \sim k^{ - \gamma }$ implies $\gamma=2(\beta+1)/\alpha$. The exponents $\alpha$ and $\beta$ can be directly measured from the data of Fig.~\ref{fig_Z233}, where we display the shape $Z(s)$ (after subtraction of $\Gamma(s)$, in order to obtain a periodic function $Z(\Gamma+1)=Z(\Gamma)$, in terms of the parameter $\Gamma$) and its spectrum together with the one of the intensity. In Fourier space we have $|Z_k|^2\sim k^{-2\alpha-2}$ and $|\kappa_k|^2\sim k^{-2\beta-2}$, it follows that $\alpha=0.22$ and $\beta=-0.66$, giving $\gamma=3$ in agreement with the exponent measured from the energy.

We found that the intensity concentrations shown in Fig.~\ref{fig_xy} are related with regions characterized by the strong advection of secondary structures. Indeed, during the evolution of secondary instabilities, filamentary sheets are advected by the previous formed intense vortices (the central tightly wounded region of the spirals). A simple model describing this situation, consists of a fixed strong point vortex advecting an initially straight weak intensity vortex sheet.

\begin{figure}
    \centering
    \includegraphics[width=0.3\textwidth]{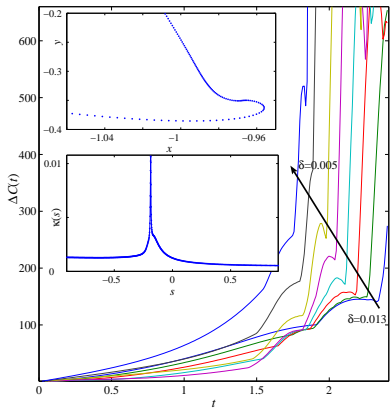}
    \caption{\label{fig_adv}(Color online.)
The maximal curvature jump as a function of time for various values of $\delta\in(0.013,0.005)$, showing the approach to a singularity in the limit $\delta\rightarrow0$, during the advection of the sheet, from numerical solutions of (\protect\ref{BRe}). Inserts: (top) Snapshot of the vortex sheet with $\delta=0.005$ at $t=1.8$; and (middle) the corresponding vortex intensity $\kappa$ as a function of the arclength $s$.
    }
\end{figure}

The evolution of the sheet $z=Z(s,t)$ of intensity $\kappa(s,t)=d\Gamma(s,t)/ds$, $s$ is the arclength, is given by a Birkhoff--Rott equation with an external velocity field due to the point vortex of circulation $\Gamma_0$:
\begin{equation}\label{BRe}
    \frac{\partial}{\partial t}\bar{Z}(s,t)=
    \frac{\Gamma_0}{2\pi\I Z(s,t)}+
    \frac{1}{2\pi\I}
    \fint\frac{\kappa(s',t) ds'}{Z(s,t)-Z(s',t)}\,.
\end{equation}
Initially one may consider that the sheet is located at a distance $d$ in the $y$ direction from the central point vortex, $Z(s,0)=s+\I d$. If one neglects the self-interaction, the advection of the sheet is described by the curve,
$Z=z_0(s,t)=(s+\I d)\E^{\I \Omega(s) t}$, where $\Omega(s) = \Gamma_0/2\pi(d^2+s^2)$. However, a singularity cannot result from the sole advection, and the effect of self-interaction is needed. Direct integration of (\ref{BRe}) actually shows that, in the limit of $\delta\rightarrow 0$, the curvature diverges at a point $s_0$, where it changes sign, leading to the eventual formation of a new spiral (Fig.~\ref{fig_adv}). In order to identify the nature of this singularity we assume that the behavior of the vortex sheet around $s=s_0$, can be analyzed within a local approximation \cite{Caflisch90}, therefore neglecting the contour integral in the computation of the principal value of the Birkhoff--Rott term,
\begin{equation}\label{BRdz}
    \frac{\partial}{\partial t}\bar{Z}(s,t)=-\frac{\I}{Z(s,t)}+\frac{a}{Z_s(s,t)}\,,
\end{equation}
where we introduced non-dimensional variables based on the time unit $2\pi d^2/\Gamma_0 $ and the space unit $d$. The non-dimensional parameter $a=\pi\kappa_0 d/\Gamma_0\ll1$ measures the circulation ratio between the central point vortex and the sheet. In the following we put the point $z=Z(s_0,t_0)$ corresponding to the singularity, at the origin $s_0=0$ and $t_0=0$. Moreover, the transformation $Z(s,t)=r(s,t)\exp[\I\phi(s,t)+\I\Omega(s) t]$, allows to separate the dynamics of the complex amplitude $r\E^{\I\phi}$ from the pure rotation at frequency $\Omega$, and leads, after substitution into (\ref{BRdz}), to the two real equations,
\begin{eqnarray}
   \frac{\partial}{\partial t} r&=&\frac{ar_s}{r_s^2+
		r^2(\phi_s+\Omega_s t)^2}\,,
        \label{r}\\
    \frac{\partial}{\partial t}\phi&=&\left(-\Omega+\frac{1}{r^2}\right)+
        a\frac{\phi_s+\Omega_s t}{r_s^2+r^2(\phi_s+\Omega_s t)^2}\,.
        \label{phi}
\end{eqnarray}
We assume now, that the form of the sheet near de singularity can be written as $r=r_0+\mathrm{sgn}(st)R)(s,t)$ and $\phi=\phi_0+\mathrm{H}(t)\Phi)(s,t)$, with the regular parts determined by the advection $r_0(s)=|z_0|=[1+(s+s_0)^2]^{1/2}$ and $\phi_0(s)=\arg(z_0)=\arctan(s+s_0)+\pi/2+\Omega t_0$, and where the singular parts $R(s,t)$ and $\Phi(s,t)$, should possess strong derivatives near $s\approx 0$. This can be satisfied if $R\sim|s|^q$, with the condition $q<1$. We introduced the sign $\mathrm{sgn}(st)$ and the Heaviside $\mathrm{H(t)}$ functions, to reproduce the jump in the curvature $|\Delta C(t)|$, as observed in the numerical computations of Fig.~\ref{fig_adv}, together with the smoothness of the phase $\Phi(s,t)$ before the singularity. Substituting these expressions into (\ref{r}-\ref{phi}), and neglecting small terms in $|s|\ll 1]$ (we can \emph{a posteriori} verify the consistency of these assumptions), we get, 
\begin{equation}\label{sing}
    \frac{\partial}{\partial t}R = a\frac{\mathrm{sgn}(s t)}{R_s}\,,\;
    \frac{\partial}{\partial t}\Phi = -\frac{\mathrm{sgn}(s)}{r_0^3}R+
        a\frac{\Phi_s+\Omega_s t}{R_s^2}\,,
\end{equation}
whose solutions, singular at $s\sim0$, are,
\begin{equation}\label{solR}
    R=(4a|st|)^{1/2}\,,
\end{equation}
and
\begin{equation}\label{solPhi}
    \Phi=\left(\Omega_s|s|\log(1/|s|)-
        \frac{\mathrm{sgn}(s)}{r_0^3}(4a|st|)^{1/2}\right)t+f(st)\,.
\end{equation}
where $f(st)$ is an arbitrary smooth function, that can be neglected. We note that the solution of (\ref{sing}) naturally leads to $q=1/2$, and to the scalings: $R_s\gg r_0\Phi_s$ for $t\ll 1$, justifying the assumption that the second term in the denominator of (\ref{r}-\ref{phi}) is negligible near the singularity. It is also worth noting that this singularity is stronger than the one found by Moore \cite{Moore79}, whose characteristic exponent is $q=3/2>1$, the difference coming from the presence of the advection velocity field.

In this Letter the $k^{-3}$ energy spectrum of two dimensional turbulence is recovered in a simple model of a single, regularized, vortex sheet evolving according to a Hamiltonian dynamics. The vortex sheet is subject to a cascade of Kelvin--Helmholtz instabilities, leading to spiral windings at different scales. The flow is characterized by an intermittent distribution of the vorticity intensity along the vortex sheet, and by a complex sequence of compression and stretching regions resulting from the nonlocal Lagrangian dynamics of the strain. The sole consideration of the fractal structure of the spiral windings is not enough to account for the observed spectrum. A simple dimensional argument using the Biot--Savart equation, allows to relate the exponents of the  singularities in the shape and the intensity of the vortex sheet, with the spectrum exponent. We propose a mechanism for the formation of strong secondary singularities, associated with jumps in the curvature, by the simultaneous action of the advection and self-interaction of the vortex sheet. A simple model describes the structure of the sheet near the singularity in the form of a square root cusp $Z(s)\sim \mathrm{sgn}(s)|s|^{q}$, with an exponent $q=1/2$. This value is sufficiently small to be in agreement with the requirement that $3<\gamma<4$. If one assumes that $q\sim\alpha$, the condition $(\beta+1)/2<q\sim\alpha<2/3$ should hold. The formation of singular concentrations of vorticity in regions with diverging curvature, constitute a dynamical mechanism towards a turbulent dissipation in the limit of vanishing viscosity.

\bibliography{toutv}

\end{document}